\documentclass{article}
\usepackage{graphicx}  % standard LaTeX graphics tool;
\usepackage{amsmath}   % For getting proper math eqns
\usepackage{amssymb}   % Used for getting various symbols eg. gtrsim, lesssim etc.
\usepackage{bm} % For bold math (esp of lower greek letters)
\usepackage{dcolumn}% Align table columns on decimal point
\usepackage{color}
\usepackage{mathrsfs}
\usepackage{amsfonts}
\usepackage{varioref}
\usepackage{mathtools}
\usepackage[toc,page]{appendix}
\RequirePackage[colorlinks,citecolor=blue,urlcolor=magenta,linkcolor=blue]{hyperref}
\labelformat{section}{Section #1} 
\labelformat{subsection}{Section #1} 
\labelformat{subsubsection}{Section #1}
\labelformat{subsubsubsection}{Section #1}
\labelformat{equation}{Eq.~(#1)} 
\labelformat{figure}{Fig.~#1} 
\labelformat{subfigure}{Fig.~\thefigure#1} 
\labelformat{table}{Tab.~#1} 
\labelformat{appendix}{Appendix #1}
\newcommand\backmatter{\appendix
\def\chaptermark##1{\markboth{%
\ifnum  \c@secnumdepth > \m@ne  \@chapapp\ \thechapter:  \fi  ##1}{%
\ifnum  \c@secnumdepth > \m@ne  \@chapapp\ \thechapter:  \fi  ##1}}%
\def\sectionmark##1{\relax}}

\title{\bf Kerr-de Sitter spacetime, Penrose process  and the generalized area theorem}
\author{\bf Sourav Bhattacharya\footnote{sbhatta@iitrpr.ac.in}\\ 
\small{Department of Physics, Indian Institute of Technology Ropar, Rupnagar, Punjab 140 001, India}}
\begin{document}
  
\maketitle
\begin{abstract}
\noindent  
We investigate various aspects of energy extraction via the Penrose process in the Kerr-de Sitter spacetime.  We show that the increase in the value of a  positive cosmological constant, $\Lambda$, always reduces the efficiency of this process.  The Kerr-de Sitter spacetime has two ergospheres -- associated with the black hole and the cosmological event horizons. We prove by analysing turning points of the trajectory that the Penrose process in the cosmological ergoregion is never possible. We next show that in this  process  both black hole and cosmological event horizons' areas increase, the later becomes possible when the particle coming from the black hole ergoregion escapes through the cosmological event horizon. We identify  a new, local mass function instead of the mass parameter, to prove this generalized  area theorem. This mass function takes care of the local spacetime energy due to the cosmological constant as well, including that arises due to the frame dragging effect due to spacetime rotation.   While the current observed value of $\Lambda$ is much tiny, its effect in this process could be considerable  in  the early universe scenario endowed with a rather high value of it, where the two horizons could have comparable sizes. In particular, the various results we obtain here are also evaluated in a triply degenerate limit of the Kerr-de Sitter spacetime we find, in which radial values of the inner, the black hole and the cosmological event horizons are nearly coincident.  
\end{abstract}

%\vskip .5cm
\noindent
%{\bf PACS : } \\
\noindent
{\bf Keywords :} Penrose process, Kerr-de Sitter, generalized area theorem, extremal limits
%\begin{document}
%\bigskip
%\maketitle
%%%%%%%%%%%

%%%%%%%%%%%%%%%%
\section{Introduction}\label{int}
%%%%%%%%%%%%%%%%
The notion of a conserved positive energy of a particle or any system is associated with a future directed timelike Killing vector field, e.g., the time translational Killing vector fields in the Minkowski or in the exterior of a static black hole spacetime. For a stationary rotating black hole, e.g. the Kerr spacetime, however, the surface at which the timelike Killing vector field becomes null has only non-zero intersections with the horizon at the axial points, $\theta=0,\pi$, where the effect of the rotation vanishes. The axisymmetric surface where the timelike Killing vector field becomes null is known as the ergosphere and the black hole event horizon (BEH) is located within it. Thus in the region between the ergosphere and the BEH, known as the ergoregion, the timelike Killing vector field is spacelike, e.g.~\cite{Wald:1984rg, Chandrasekhar:1985kt, Paddy}. Due to this non-existence of a future directed timelike Killing vector field,  there can be negative energy particles within the ergoregion, even classically.

The existence of such negative energy particles gives rise to a classical mechanism for energy extraction from a black hole, namely the Penrose process, as follows~\cite{Wald:1984rg, Chandrasekhar:1985kt,  Paddy}, also references therein. Let a particle carrying positive energy enters the ergoregion of a Kerr black hole. We imagine that it breaks into two pieces there -- one carrying negative energy and the other, positive. The negative energy particle enters the BEH and the positive energy particle, after reaching a turning point, comes out of the ergosphere and finally gets intercepted by an outside observer. With respect to such an observer the usual (positive) energy conservation must be valid. Thus it is clear that the ejecta will carry more energy than the initial incoming particle,  effectively extracting energy from the black hole. It turns out that the energy thus extracted is largely rotational and the process can only continue until the black hole settles down into the  Schwarzschild spacetime, where no ergosphere is present.

Since the Penrose process reduces more the rotation of a black hole than its mass, a Kerr black hole can never become a naked curvature singularity under this process, see~\cite{Wald} for a formal proof of this. In~\cite{Christodoulou:1970wf}, it was proved that the black hole horizon area must increase under this process,
thereby providing evidence in favour of the second law of black hole mechanics. We further refer our reader to~\cite{Wald:1974kya, Bardeen:1972fi} for various inequalities 
(respectively, the Wald and the Bardeen-Press-Teukolsky inequality) regarding the local speed of the fragments within the ergoregion. In order that the process indeed  occurs, those inequalities show that the  fragments must be moving with considerable relativistic speed. We further refer our reader to~\cite{Christodoulou:1972kt} for an interesting account on how electrostatic energy could be extracted from a charged black hole, via a Penrose-like process. It is also relevant to note that if we consider a magnetic field associated with the black hole and compute the Penrose process for a test charge particle, the energy extraction turns out to be much more efficient than the uncharged case~\cite{Wagh:1986tsa, Dadhich:2012yu}.

A variant of the Penrose process using classical fields instead of particles exists, namely, the superradiance or the superradiant scattering, e.g.~\cite{Wald:1984rg}. If $\omega$ is the frequency of a scalar field and $\Omega_H$ is the angular speed on the horizon, for $(\omega-\lambda \Omega_H) <0$ ($\lambda$ denotes the azimuthal eigenvalue), the flux of the energy momentum going through the horizon turns out to be negative whereas the flux at infinity is positive. It thus effectively extracts energy from the hole. We refer our reader to e.g.~\cite{Leite:2017zyb}-\cite{Zhu:2014sya}  for most recent trends and developments in this topic. See also \cite{Ganguly:2017zjf} and references therein for a discussion of superradiance in the acoustic analogue gravity paradigm.
We further refer out reader to~\cite{Brito:2015oca} for a review and an exhaustive list of references. In particular, superradiance has  been studied for the anti-de Sitter black hole spacetimes~\cite{Winstanley:2001nx, Aliev:2015wla, Bosch:2016vcp, Gonzalez:2017shu}, including nonlinear backreaction effects~\cite{Bosch:2016vcp}. For de Sitter black holes on the other hand, which is our focus, we refer our reader to~\cite{Tachizawa:1992ue, Anninos:2010gh, Zhang:2014kna, Georgescu:2014yea, Zhu:2014sya}. The condition for energy extraction for such black holes is given by $(\omega-\lambda \Omega_C)>0$ where $\Omega_C$ is the angular speed on the cosmological event horizon, along with the usual $(\omega-\lambda \Omega_H) <0$.   In particular, massless scalar wave superradiance was studied in~\cite{Anninos:2010gh} in the doubly degenerate Nariai limit of the Kerr-de Sitter spacetime. 

In this paper, we shall study the Penrose process in the Kerr-de Sitter (KdS) spacetime~\cite{Carter:1968ks, Gibbons:2004uw, Akcay:2010vt, Lake:2015xca},  and would address a couple of questions which, to the best our knowledge, has  been hitherto unaddressed in the literature. Firstly, it is relevant to ask,  how does a positive $\Lambda$
affect the process' efficiency, keeping in mind the results of~\cite{Wald:1974kya, Bardeen:1972fi} pertaining the asymptotically flat spacetime? We shall see that in  KdS this process becomes {\it always} less efficient with increasing $\Lambda$ (cf. \ref{sec.3}, \ref{sec.4}).  Second, unlike the Kerr, KdS is endowed with two ergospheres (cf. discussions of \ref{sec.2}), associated with the black hole and the cosmological event horizon (CEH). It is natural to ask then, can there be any Penrose process in the cosmological ergosphere as well? Clearly, in such process one expects to steal rotational energy  from regions beyond the CEH, induced onto the the cosmological constant due to frame dragging effects (cf. \ref{sec.3.1}). However,  we shall see by a simple analysis of the turning points of the trajectory  that such process is never possible.

The third question would concern the second law of the de Sitter black hole thermodynamics,  see e.g.~\cite{Gibbons:1977mu}-\cite{Bhattacharya:2017bpl} and references therein for recent developments. Precisely, the two horizons (BEH and CEH) in such spacetimes have individual temperature and entropy associated with them~\cite{Gibbons:1977mu}. A rigorous proof of the area theorem for the cosmological event horizon can be seen in~\cite{Maeda:1997fh}. Clearly, such two temperature system makes the usual notion of the black hole thermodynamics as a grand canonical ensemble ambiguous~\cite{Traschen:1999zr}.  To an observer residing within the region bounded by the two horizons,   a total entropy of the spacetime can be attributed, equaling the quarter of the sum of the two horizon areas~\cite{Davies:2003me} (see also e.g.~\cite{Urano:2009xn, Saida:2009up, Bhattacharya:2013tq}  and references therein). We refer our reader to~\cite{Bhattacharya:2015mja} for a derivation of such total entropy using near horizons' conformal symmetries. We also refer our reader to~\cite{Li:2016zdi} for a proposal of further modifying such entropy via correlation between the two horizons. Quite interestingly, the variation of the  total entropy gives rise to a {\it single} effective temperature in such spacetimes, e.g.~\cite{Urano:2009xn, Saida:2009up, Bhattacharya:2013tq}. Interesting results, including  calculations pertaining phase transition using such effective temperature can be seen e.g. in~\cite{Zhang:2014jfa, Kubiznak:2015bya, Kanti:2017ubd, Pappas:2017kam}. It is thus a very  important task to check whether actually in a physical scenario the second law of de Sitter  black hole mechanics holds i.e., whether in such process the aforementioned total entropy increases, known as the generalized area theorem~\cite{Davies:2003me}. There has been attempt to check this theorem under different physical processes, however, to the best of our knowledge, unlike the asymptotically flat spacetimes, a completely general or topological proof of this theorem is hitherto absent.

Motivated by this,  in \ref{sec.5}, we shall prove by introducing a suitable local mass function that in the Penrose process, not only the area of the BEH increases, but also when the ejecta escapes through the CEH, its area gets increased too, thereby proving the generalized area theorem in this process. Finally we conclude in \ref{sec.6}. We shall use mostly positive signature for the metric and will set $c=1=G$ below.

%%%%%%%%%%%%%%%%%%%%%%%%%%%%%%%%%%%%%%%%%%%%%%%%%%%%%%%%%%%%%%%%%%%%%%%%%%
\section{The metric and some general feature of horizon sizes}\label{sec.2}
%%%%%%%%%%%%%%%
The Kerr-de Sitter (KdS) metric in the Boyer-Lindquist coordinates reads~\cite{Carter:1968ks, Gibbons:2004uw, Akcay:2010vt, Lake:2015xca},
\begin{eqnarray}
ds^2=-\frac{\Delta_r-a^2\sin^2\theta \Delta_{\theta}}{\rho^2}dt^2 -\frac{2a \sin^2 \theta }{\rho^2 \Xi} \left( (r^2+a^2 )\Delta_{\theta}-\Delta_r\right)dt d\phi \nonumber\\ + \frac{\sin^2 \theta }{\rho^2 \Xi^2} \left( (r^2 +a^2)^2 \Delta_{\theta}-\Delta_r a^2 \sin^2\theta\right)d\phi^2 + \frac{\rho^2}{\Delta_r}dr^2 + \frac{\rho^2}{\Delta_{\theta}}d\theta^2
\label{sup1}
\end{eqnarray}
where,
\begin{eqnarray}
\Delta_r = (r^2 +a^2) (1-H_0^2 r^2 ) -2Mr, \quad \Delta_{\theta} =1 + H_0^2 a^2 \cos^2\theta,  \quad \Xi = 1+ H_0^2 a^2,  \quad \rho^2 =r^2 +a^2 \cos^2 \theta 
\label{sup2}
\end{eqnarray}
where $H_0^2 =\Lambda/3$ with $\Lambda$ being the positive cosmological constant and $M$ and $a$ are usually respectively called the mass and angular momentum parameters. Setting $a=0$ recovers the Schwarzschild-de Sitter spacetime whereas setting further $M=0$ recovers the de Sitter spacetime written in the static patch. Setting $M=0$ alone results in a line element diffeomorphic to the de Sitter~\cite{Akcay:2010vt}.  The cosmological and the black hole event horizons are respectively given by the largest (say, $r_C$) and the next to the largest (say, $r_H$) roots of $\Delta_r=0$. The smallest positive root, $r=r_-$ of $\Delta_r=0$ corresponds to the inner or the Cauchy horizon.

The location of the ergosphere for a stationary axisymmetric spacetime is given by $g_{tt}=0$~\cite{Wald:1984rg}. Unlike the $\Lambda\leq 0$ cases, the KdS spacetime is endowed with two, instead of one ergosphere as follows. Since $g_{tt}>0$ at $\Delta_r =0$, the Killing vector field $(\partial_t)^a$ is spacelike on the horizons. Thus $(\partial_t)^a$ must be null at some points off the horizons. Since there exists region in between the black hole and the cosmological event horizon where $(\partial_t)^a$ is timelike, it is clear that we must have two surfaces on which $g_{tt}=0$, giving two ergospheres associated with the two Killing horizons. We shall call them the black hole ergosphere and the cosmological ergosphere and will  investigate the Penrose process  for both of them.

%The inverse metric functions are given by
%
%\begin{eqnarray}
%&&g^{tt}=-\frac{(r^2+a^2)^2\Delta_{\theta}-\Delta_r a^2\sin^2\theta}{\Delta_r \,\Delta_{\theta} \,\rho^2}, \qquad g^{t\phi}=-\frac{\Xi a \left((r^2+a^2)\Delta_{\theta}-\Delta_r\right)}{\Delta_r\, \Delta_{\theta}\, \rho^2} \nonumber\\ &&g^{\phi\phi}=\frac{\Xi^2 \left(\Delta_{r}-a^2\sin^2\theta\Delta_{\theta}\right)}{\Delta_r\, \Delta_{\theta}\, \rho^2 \sin^2\theta}, \qquad  
%g^{rr}= \frac{\Delta_r}{\rho^2}, \qquad g^{\theta\theta}=\frac{\Delta_{\theta}}{\rho^2}
%\label{sup3}
%\end{eqnarray}
%
The surface gravities of the black hole event horizon (BEH) and the cosmological event horizon (CEH)  of \ref{sup1} are respectively given by
\begin{eqnarray}
\kappa_{H}= \frac{r_H\left(1-2H_0^2 r_H^2-H_o^2 a^2\right)-M }{r_H^2+a^2}\qquad -\kappa_C= \frac{r_C\left(1-2H_0^2 r_C^2-H_o^2 a^2\right)-M }{r_C^2+a^2}
\label{sup10}
\end{eqnarray}
where $\kappa_C$ is positive and the minus sign in front of it indicates repulsive effects. The areas of the BEH and CEH are respectively given by
\begin{eqnarray}
A_{H}= \frac{4\pi (r_H^2+a^2)}{1+H_0^2a^2}\qquad A_C= \frac{4\pi (r_C^2+a^2)}{1+H_0^2a^2}
\label{sup10'}
\end{eqnarray}
We have the angular speeds $(=-g_{t\phi}/g_{\phi\phi})$ on the horizons,
\begin{eqnarray}
\Omega_{H}=\frac{a\Xi }{r_H^2+a^2} \qquad \Omega_{C}=\frac{a\Xi }{r_C^2+a^2} 
\label{sup10s}
\end{eqnarray}
and also the horizon Killing vector fields,
\begin{eqnarray}
\chi_H^a=(\partial_t)^a+\Omega_H (\partial_{\phi})^a \qquad \chi_C^a=(\partial_t)^a+\Omega_C (\partial_{\phi})^a, 
\label{sup10k}
\end{eqnarray}
future directed and null on the respective horizons.\\

\noindent
We further note below some general features regarding the comparative horizons' radii of KdS, the Kerr, the Schwarzschild-de Sitter (SdS) and the empty de Sitter, useful for our future purpose, derived in \ref{A} : a) the CEH of the KdS is smaller than the empty de Sitter radius, $H_0^{-1}$. b) the BEH of KdS is smaller than that of  SdS but CEH of KdS is larger than the CEH of SdS (the parameter $M$ is held fixed). c) the BEH of KdS is larger than the BEH of the Kerr, with the parameters $M$ and $a$ held fixed. Since the minimum horizon length 
of the Kerr is $M$ or $a$ (the extremal case), clearly for the KdS we have a lower bound on the BEH radius, $r_H>M$ or $r_H>a$. Also, as we increase $H_0$, the BEH radius of KdS increases whereas the CEH radius decreases.  All these conclusions on  horizon  sizes in \ref{A} is derived  by playing with the sign of the function $\Delta_r$ (\ref{sup2}) only, in different spacetime regions. 

We also note here the limits of the various parameters for a regular KdS spacetime derived in \ref{B}, starting from a triply degenerate limit in which the inner horizon, the BEH and CEH coincide, 
\begin{eqnarray}
MH_0 \leq 0.2435 ~~{\rm and}~~\frac{a}{M} \leq 1.01 \,  ~~{\rm and}~~ aH_0 \leq 0.246
\label{a7'}
\end{eqnarray}
Clearly only any two of the above inequalities can be independent. 
Note that the upper bound on $aH_0$ matches with the one found numerically in~\cite{Anninos:2010gh}. Note also that $a=M$ is not the extremal limit of 
KdS, pointed out earlier by~\cite{Lake:2015xca} via analyzing the parameter space. 

%Below we shall consider the variation of $M$ and $a$ only and treat $\Lambda$ to be a constant.

%%%%%%%%%%%%%%%%%
\section{Calculation of the Penrose process }\label{sec.3}
%%%%%%%%%%%%%%%%%%%%%%%%%%%%%%%%%%%%
\noindent 
We shall briefly sketch below the derivation of the variable separated geodesic equation for the KdS spacetime, first done in~\cite{Carter:1968ks}. The stationarity and axisymmetry of the KdS spacetime permits two conserved quantities for a geodesic $u^a$ -- the energy and the orbital angular momentum, respectively given by
\begin{eqnarray}
E=-g_{ab}(\partial_t)^{a}u^{b} \qquad  L= g_{ab}(\partial_{\phi})^{a}u^{b}
\label{sup3'}
\end{eqnarray}
In order to analyze geodesics and make  variable separation in \ref{sup1},  the Hamilton-Jacobi equation is useful~\cite{Chandrasekhar:1985kt, Carter:1968ks}, 
\begin{eqnarray}
2\,\frac{\partial S}{\partial \lambda}=g^{ab}\left(\frac{\partial S}{\partial x^a}\right)\left(\frac{\partial S}{\partial x^b}\right)
\label{sup4}
\end{eqnarray}
where $\lambda$ is an affine parameter along the geodesic and we take the ansatz for Hamilton's principle function $S$ as
\begin{eqnarray}
S=-\frac12 \delta \lambda -E t+L\phi + S_r(r) + S_{\theta}(\theta) 
\label{sup5}
\end{eqnarray}
where $\delta=1\,(0)$ for timelike (null) geodesics. \ref{sup4} can be written explicitly as
\begin{eqnarray}
&& \rho^2 \delta  = \frac{(r^2+a^2)^2\Delta_{\theta}-\Delta_r a^2\sin^2\theta}{\Delta_r\,\Delta_{\theta}}\,E^2-\frac{2\,\Xi a \left((r^2+a^2)\Delta_{\theta}-\Delta_r\right)}{\Delta_r\, \Delta_{\theta}}E L -\frac{\Xi^2 \left(\Delta_{r}-a^2\sin^2\theta\Delta_{\theta}\right)}{\Delta_r\, \Delta_{\theta}\, \sin^2\theta}L^2\nonumber\\
&&-\Delta_{\theta} \left(\frac{\partial S_{\theta}(\theta)}{\partial \theta}\right)^2 -\Delta_r \left(\frac{\partial S_{r}(r)}{\partial r}\right)^2 
\label{sup6}
\end{eqnarray}
which, after a little  rearrangement could be written as 
\begin{eqnarray}
&&\Delta_r \left(\frac{\partial S_{r}(r)}{\partial r}\right)^2-\frac{(r^2+a^2)^2}{\Delta_r}\left(E-\frac{a\Xi L}{r^2+a^2}\right)^2+(\Xi L-aE)^2 +\delta r^2\nonumber\\ &&=
-\Delta_{\theta} \left(\frac{\partial S_{\theta}(\theta)}{\partial \theta}\right)^2-\frac{\sin^2\theta}{\Delta_{\theta}}\left(aE-\frac{\Xi L}{\sin^2\theta}\right)^2+(\Xi L-aE)^2 -\delta a^2\cos^2\theta
\label{sup7}
\end{eqnarray}
Thus the left\,(right)  hand side of the above equation is a function of $r\,(\theta)$ only. This is only possible if each side equals a constant, known as the Carter constant, say, $-\zeta$. Also, the constant $(\Xi L-aE)^2 $ appearing on both the sides guarantee the recovery of correct equation in the static limit ($a=0$). A proof that $\zeta \geq 0$ is given in \ref{C}. 
The momenta along the radial and polar directions are given by
\begin{eqnarray}
p_r=\frac{\partial S_r}{\partial r} \qquad p_{\theta}=\frac{\partial S_{\theta}}{\partial \theta }
\label{sup8}
\end{eqnarray}
Thus, $p^{r}\equiv\dot{r}= g^{rr}p_r$ and $p^{\theta}\equiv \dot{\theta}=g^{\theta \theta}p_{\theta}$, giving us the radial and polar equations for the geodesic, 
\begin{eqnarray}
&&\rho^4\dot{r}^2=\left(r^2+a^2\right)^2\left(E-\frac{a\Xi L}{r^2+a^2}\right)^2-\Delta_r\left(\zeta +\delta r^2+(\Xi L -aE)^2\right) \nonumber\\ 
&& \rho^4 \dot{\theta}^2= -\frac{1}{\sin^2\theta}\left(Ea\sin^2\theta-\Xi L\right)^2+\Delta_{\theta}\left(\zeta -\delta a^2 \cos^2\theta +(\Xi L -aE)^2 \right)
\label{sup9}
\end{eqnarray}

\noindent 
Let us now derive the Penrose process following~\cite{Chandrasekhar:1985kt}.    Imagine that one massive particle moving along a timelike geodesic enters the ergosphere and reaches one turning point ($\dot{r}=0$) and there it breaks into two massless pieces -- with negative and positive energies. For  $\dot{r}=0$, we obtain from the first of \ref{sup9},
\begin{eqnarray}
E=\frac{\Xi aLx \pm \left[\Xi^2 a^2L^2x^2+ \left(r^2+a^2(1+x)\right)\left(\Xi^2L^2(1-x)+\Delta_r(\delta+\zeta/r^2)\right)\right]^{1/2}}{\left[r^2+a^2(1+x)\right]}
\label{sup11}
\end{eqnarray}
or alternatively,
\begin{eqnarray}
L=\frac{-aEx \pm \left[ a^2E^2x^2+ (1-x) \left(E^2\left(r^2+a^2(1+x)\right)-\Delta_r(\delta+\zeta/r^2) \right)\right]^{1/2}}{\Xi(1-x)}
\label{sup12}
\end{eqnarray}
where for the sake of brevity we have written,
\begin{eqnarray}
x=H_0^2(r^2+a^2)+\frac{2M}{r}
\label{sup12'}
\end{eqnarray}
Let us now consider a null geodesic  ($\delta=0$) on the equatorial plane $\theta =\pi/2$ (i.e., $\zeta=0$) and consider \ref{sup11} first, where, in order to have positive energy in the limit $a \to 0$, we must retain only the positive sign. On the other hand  for $a\neq 0$, a necessary  criterion for having negative energy is $L<0$ (i.e. counter rotating orbits with respect to the direction of the black hole angular momentum).  Then it is clear that along with the necessary condition $L<0$, in order to have negative energy we must also have
\begin{eqnarray}
1-2M/r-H_0^2(r^2+a^2) <0
\label{sup13}
\end{eqnarray}
Since $g_{tt}(\theta=\pi/2)= -\left(1-H_0^2(r^2+a^2)-2M/r\right)$, the above inequality clearly represents the inside-ergosphere region on the equatorial plane.  A little algebra simplifies \ref{sup12} to
\begin{eqnarray}
L=\frac{E\left[-a\left(\frac{2M}{r}+H_0^2(r^2+a^2)\right)\pm \sqrt{\Delta_r} \left(1-\frac{\delta\left(1-\frac{2M}{r}-H_0^2(r^2+a^2)\right)}{E^2} \right)^{1/2}\right]}{\Xi\left(1-\frac{2M}{r}-H_0^2(r^2+a^2)\right)}
\label{sup14}
\end{eqnarray}
Let us now consider a massive particle ($\delta=1$) with energy $E^{(0)}>0$ and angular momentum $L^{(0)}$ entering the ergosphere. Let the particle be broken into two massless ($\delta=0$) pieces with energies and angular momenta $(E^{(1)}, L^{(1)})$ and $(E^{(2)}, L^{(2)})$ respectively, 
one crossing the black hole horizon while the other coming out of the ergoregion. We have from \ref{sup14}
\begin{eqnarray}
&&L^{(0)}= E^{(0)}\frac{\left[-a\left(\frac{2M}{r}+H_0^2(r^2+a^2)\right)+ \sqrt{\Delta_r} \left(1-\frac{\left(1-\frac{2M}{r}-H_0^2(r^2+a^2)\right)}{(E^{(0)})^2} \right)^{1/2}\right]}{\Xi\left(1-\frac{2M}{r}-H_0^2(r^2+a^2)\right)} = \alpha^{(0)}(r,E^{(0)}) E^{(0)} ~~~(\rm say)\nonumber\\
&&L^{(1)}= E^{(1)}\frac{\left[-a\left(\frac{2M}{r}+H_0^2(r^2+a^2)\right) -\sqrt{\Delta_r}\right]}{\Xi\left(1-\frac{2M}{r}-H_0^2(r^2+a^2)\right)}= \alpha^{(1)}(r) E^{(1)} ~~~(\rm say)\nonumber\\
&&L^{(2)}= E^{(2)}\frac{\left[-a\left(\frac{2M}{r}+H_0^2(r^2+a^2)\right)+ \sqrt{\Delta_r}\right]}{\Xi\left(1-\frac{2M}{r}-H_0^2(r^2+a^2)\right)}= \alpha^{(2)}(r) E^{(2)} ~~~(\rm say)
\label{sup15}
\end{eqnarray}
Using the above equations, the stationary-axisymmetric conservation laws  read
\begin{eqnarray}
E^{(1)}+E^{(2)}=E^{(0)} \qquad L^{(1)}+L^{(2)}= \alpha^{(1)}(r)E^{(1)}+\alpha^{(2)}(r)E^{(2)} =L^{(0)}= \alpha^{(0)}(r,E^{(0)}) E^{(0)} 
\label{sup16}
\end{eqnarray}
which can be solved to get the energies
\begin{eqnarray}
E^{(1)}=\frac{\alpha^{(0)}(r, E^{(0)})-\alpha^{(2)}(r)}{\alpha^{(1)}(r)-\alpha^{(2)}(r)} E^{(0)} \qquad E^{(2)}=\frac{\alpha^{(1)}(r)-\alpha^{(0)}(r,E^{(0)})}{\alpha^{(1)}(r)-\alpha^{(2)}(r)}E^{(0)}
\label{sup17}
\end{eqnarray}
We find using \ref{sup15},
\begin{eqnarray}
E^{(1)}=-\frac12 \left[ \left(1-\frac{\left(1-\frac{2M}{r}-H_0^2(r^2+a^2)\right)}{(E^{(0)})^2} \right)^{1/2}-1\right]E^{(0)}, \quad
E^{(2)}=\frac12 \left[ \left(1-\frac{\left(1-\frac{2M}{r}-H_0^2(r^2+a^2)\right)}{(E^{(0)})^2} \right)^{1/2}+1\right]E^{(0)} \nonumber\\
\label{sup18}
\end{eqnarray}
Thus it is clear once again that energy extraction i.e., $E^{(2)}>E^{(0)}$ would be possible  when we are inside the ergosphere, \ref{sup13}. The amount of energy extracted is given by 
\begin{eqnarray}
\delta E = -E^{(1)}
\label{sup19}
\end{eqnarray}
Thus the maximum of the energy extracted would correspond to the minimum (negative) value of the function  $(1-2M/r-H_0^2(r^2+a^2))$, which certainly corresponds to  the horizon, $\Delta_r=0$ and we have
\begin{eqnarray}
\delta E_{\rm max}= \frac{1}{2}\left[\left(1+\frac{a^2}{r_{H}^2 (E^{(0)})^2}\right)^{1/2}-1\right]E^{(0)}
\label{sup20}
\end{eqnarray}
The above expression is formally similar for black hole spacetimes with or without a $\Lambda$, the effect of which implicitly comes through the value of $r_H$. For the extremal Kerr black hole $(a=M, ~H_0=0)$, one has $r_H=M$ and $1+a^2/r_H^2=2$. Taking $E^{(0)}=1$, we get $\delta E_{\rm max}= 0.207$~\cite{Chandrasekhar:1985kt}.

For the KdS, as we have seen that $r_H$ increases as we increase $H_0$, \ref{sec.2}, we conclude that a positive $\Lambda$ {\it always} reduces the efficiency of the Penrose process, for given values of $a$ and $E^{(0)}$.  In particular as a special case, if we consider the triply degenerate  limit of the KdS, \ref{B},  we get using \ref{a6} and \ref{a7},
\begin{eqnarray}
\delta E_{\rm max}= \frac{1}{2}\left[\left(1+\frac{0.469}{(E^{(0)})^2}\right)^{1/2}-1\right]E^{(0)}
\label{sup21}
\end{eqnarray}
For the customary value $E^{(0)}=1$, we find $\delta E_{\rm max}= 0.106$ which is half of the result for the extremal Kerr.  We shall further come back to this issue once again in \ref{sec.4}.

%%%%%%%%%%%%%%%%%%%%%%%%%%%%%%
\subsection{Is energy extraction using the cosmological ergosphere possible?}\label{sec.3.1}
%%%%%%%%%%%%%%%%%%%%
Let us imagine a particle carrying positive energy and  angular momentum breaks into two fragments in the {\it cosmological ergoregion}. One of the fragments, carrying negative energy and angular momentum crosses the CEH and escapes while the other carrying positive energy and angular momentum comes into the region inside. The rotation of the black hole induces  frame dragging effect onto $\Lambda$, as is evident from the $H_0^2a^2$ term appearing in various metric functions. Thus in the cosmological ergoregion, the Penrose process, if any, is supposed to steal rotational kinetic energy from the cosmological constant, in the region beyond CEH. However, such process is never possible, as can be seen below.

Firstly for the black hole, it is clear that in order to ensure that the particle carrying positive energy and angular momentum indeed comes out of the ergosphere instead of falling into the hole, we must have a turning point, say $r=r_T$, somewhere in between the ergospehere and the horizon along its trajectory~\cite{Chandrasekhar:1985kt}\footnote{We have seen in the preceding discussions that the maximum amount of energy is extracted when the turning point is located on the horizon.}. Let us consider the first of \ref{sup9} with $\dot{r}^2\,(r=r_{T}+\delta r_T;\,\delta r_T>0)=0^+$. Thus if we move inward, for $E,\,L>0$, we must have $\dot{r}^2(r=r_T) = 0$, for $r_T$ as a turning point to exist. Since both $\Delta_r$ and $(E-a\Xi L/(r^2+a^2))^2$  decreases with decreasing $r$, clearly it is possible  to  have such turning points. In the cosmological ergoregion on the other hand, for $E,\,L>0$, let us imagine a point where  $\dot{r}^2(r=r_T-\delta r_T;\,\delta r_T>0) = 0^{+}$. Since the function $(E-a\Xi L/(r^2+a^2))$  increases with increasing $r$, whereas $\Delta_r$ decreases,
we must have $\dot{r}^2(r=r_T) > 0^{+}$, always. This shows that there can be no such turning point and both positive and negative energy fragments, if they are created in the cosmological ergoregion, would eventually cross the CEH and disappear.

Thus  only in the black hole ergoregion the  energy extraction via the Penrose process is possible. Also, after extracting energy from the black hole, the positive energy ejecta  can reach CEH and escape, eventually decreasing the energy and angular momentum of the spacetime region bounded by CEH.
%
%\begin{figure}[h]
%\begin{center}
%\includegraphics[width=3.0 in, height=4.0 in, angle=0]{sourav.eps}
%\caption{
%Schematic representation of the Penrose process in Kerr-de Sitter, in the region bounded by the future and the  past  black hole and cosmological horizons ($H^{\pm},\,C^{\pm}$) on $\theta=\pi/2$. The analytically continued regions beyond the horizons are not shown. The dashed line denotes the ergospheres' radii on the equatorial plane. Trajectory A denotes energy extraction from the black hole and the ejecta finally moving out of the future CEH. Trajectory B just shows if any positive and negative energy pair is created in the cosmological ergoregion,  both must escape through $C^+$. 
%}
%\end{center}
%\label{fig.1}
%\end{figure}
%
%%%%%%%%%%%%%%%%%%%%%%%%%%%%%%%%%
\section{Inequalities for the local speeds of fragments }\label{sec.4}
%%%%%%%%%%%%%%%%%%%%%%%%%%%%%%%
So far we have seen that in order to do the energy extraction, we must have $L<0$. The Wald inequality~\cite{Wald:1974kya}  and the Bardeen-Press-Teukolsky inequality~\cite{Bardeen:1972fi} (see also~\cite{Chandrasekhar:1985kt}) further establishes lower bounds on ejecta particles' speed, in order that Penrose process indeed occurs.  We shall consider these inequalities in the Kerr-de Sitter spacetime below, in order to show that a positive $\Lambda$ increases those lower bounds. The derivations presented below parallel to that of the Kerr geometry and hence we shall not go into much detail, referring the reader to the above references for the same.

\subsection{The Wald inequality}
%%%%%%%%%%%%%%%%%%%%%%%%%
Let us imagine a test particle moving along a geodesic with four velocity $u^a$ ($u\cdot u=-1$) and conserved energy $E>0$ and it breaks into two fragments and let $v^a$ ($v\cdot v=-1$) and $\epsilon$  respectively be the four velocity and conserved energy of one of them. Let us erect an orthonormal basis $e_{(\mu)}{}^a$ (the Greek index within parenthesis represent the local Lorentz frame) and let $u^a=e_{(0)}{}^a$. Expanding $v^a$ then in the orthonormal basis : $v^a=e_{(\mu)}{}^a u^{(\mu)}$, we have  
\begin{eqnarray}
v^a = \frac{ \left(u^a+v^{(i)}e_{(i)}{}^a\right)}{\sqrt{1- v^{(i)}v_{(i)} }} \qquad  (i=1,2,3)
\label{sup22}
\end{eqnarray}
We next expand the timelike Killing vector field $(\partial_t)^a$ in the orthonormal basis,
\begin{eqnarray}
(\partial_t)^a=(\partial_t)^{(0)}u^a+e_{(i)}{}^a(\partial_t)^{(i)}
\label{sup23}
\end{eqnarray}
Thus the conserved energy of the initial particle, $E=-g_{ab}u^a(\partial_t)^b$, can be written in the orthonormal basis as   
\begin{eqnarray}
E= (\partial_t)^{(0)}
\label{sup24}
\end{eqnarray}
Also
\begin{eqnarray}
g_{tt}=g_{ab}(\partial_t)^a(\partial_t)^b=-((\partial_t)^{(0)})^2+ (\partial_t)^{(i)}(\partial_t)_{(i)}=-E^2+(\partial_t)^{(i)}(\partial_t)_{(i)}
\label{sup25}
\end{eqnarray}
Likewise the energy $\epsilon$ of one the fragments is given by
\begin{eqnarray}
\epsilon=-(\partial_t)^av_a= \frac{E-{\bf| v| \,|\partial_t|}\cos \varphi   }{ \sqrt{1- {\bf |v|}^2} }
\label{sup26}
\end{eqnarray}
where we have written ${\bf| v|}$ and ${\bf| \partial_t|}$ respectively as the norms of the spatial parts of the four velocity and the timelike Killing vector field  in the orthonormal frame and $\varphi$ is the angle between them. Using \ref{sup25} and \ref{sup1} the above equation becomes
\begin{eqnarray}
\epsilon= \frac{E-{\bf| v|} \left(E^2-\frac{\Delta_r-a^2\sin^2\theta \Delta_{\theta}}{\rho^2}\right)^{1/2}\cos \varphi   }{ \sqrt{1- 
{\bf |v|}^2 } }
\label{sup27}
\end{eqnarray}
This gives the Wald inequality for the Kerr-de Sitter spacetime
\begin{eqnarray}
\frac{E-{\bf| v|} \left(E^2-\frac{\Delta_r-a^2\sin^2\theta \Delta_{\theta}}{\rho^2}\right)^{1/2}   }{ \sqrt{1- {\bf |v|}^2} }\leq \epsilon \leq \frac{E+{\bf| v|} \left(E^2-\frac{\Delta_r-a^2\sin^2\theta \Delta_{\theta}}{\rho^2}\right)^{1/2}   }{ \sqrt{1- {\bf |v|}^2 } } 
\label{sup28}
\end{eqnarray}
Thus in order to have a negative value of $\epsilon$, we must have 
\begin{eqnarray}
{\bf| v|}> \frac{E}{\left(E^2-\frac{\Delta_r-a^2\sin^2\theta \Delta_{\theta}}{\rho^2}\right)^{1/2}}
\label{sup29}
\end{eqnarray}
It is clear that the above lower bound has a minimum at $\theta=\pi/2$ and on the BEH
\begin{eqnarray}
{\bf| v|}> \frac{E}{\left(E^2+\frac{a^2}{r_{H}^2}\right)^{1/2}}
\label{sup30}
\end{eqnarray}
Since $r_H$ increases with increasing $H_0$, the above expression shows that ${\bf| v|}$ would be higher with higher values of $H_0$ ($E,~a$ held fixed).
We can compare  the extreme cases here as well. For an extreme Kerr black hole, we have $a/r_H=1$. On the other hand, for the triply degenerate KdS solution, using \ref{a5}, \ref{a5'} and \ref{a6}, we find $(a/r_H)^2=(2\sqrt{3}-3)\approx 0.464$.  Thus taking the customary value  $E=1$, we find for the extreme Kerr spacetime, ${\bf| v|}>0.707$ whereas in our triply degenerate limit, we obtain ${\bf| v|}>0.825$.

%%%%%%%%%%%%%%
\subsection{The Bardeen-Press-Teukolsky inequality}
%%%%%%%%%%
The Bardeen-Press-Teukolsky inequality establishes result analogous to above  as follows. Let us consider two particles with energies $E_{+}$ and $E_{-}$ collide at a point. We erect an orthonormal basis $e_{(\mu)}{}^a$ as earlier (with $e_{(0)}=u^a$). Let us also suppose that in this frame the particles move with equal and opposite 
three velocities, $v^{(i)}$ and $-v^{(i)}$. The magnitude of their local relative speed in this frame is found by the velocity addition formula,
\begin{eqnarray}
{\bf| v_{\rm rel}|}= \frac{2 {\bf |v|}}{1+{\bf |v|}^2}
\label{sup31}
\end{eqnarray}
We wish to find a lower bound on ${\bf| v_{\rm rel}|}$ such that $E_{-}$ could be negative. Using steps similar to the previous subsection, one gets
\begin{eqnarray}
E_{\pm}=\frac{(\partial_t)^{(0)}\pm {\bf| v| \,|\partial_t|}\cos \varphi   }{ \sqrt{1- {\bf |v|}^2} }
\label{sup32}
\end{eqnarray}
Using the above expression, we find after some algebra, 
\begin{eqnarray}
{\bf |v|}^4\left(E_{+}+E_{-}\right)^2-2{\bf |v|}^2\left(E_{+}^2+E_{-}^2+2g_{tt}\right)+(E_{+}-E_{-})^2\leq 0
\label{sup33}
\end{eqnarray}
which yields
\begin{eqnarray}
{\bf |v|}\geq \frac{\sqrt{E_{+}^2-\frac{\Delta_r-a^2\sin^2\theta \Delta_{\theta}}{\rho^2}}-\sqrt{E_{-}^2-\frac{\Delta_r-a^2\sin^2\theta \Delta_{\theta}}{\rho^2}} }{\left(E_{+}+E_{-}\right)}
\label{sup34}
\end{eqnarray}
Considering the marginal  case, $E_{-}=0^-$, we find on the horizon for $\theta=\pi/2$,
\begin{eqnarray}
{\bf |v|}\geq \frac{1}{E_+}\left(\sqrt{E_+^2+\frac{a^2}{r_{H}^2}}-\frac{a}{r_{H}} \right)
\label{sup35}
\end{eqnarray}
Partially differentiating this with respect to $r_H$, we find
$$ \frac{\partial |{\bf v}|}{\partial r_H}\geq \frac{a\left(1-\frac{1}{(1+r_H^2E_+^2/a^2)^{1/2}} \right)}{r_H^2E_+}\geq 0$$
This shows as earlier that, since the increase in $H_0$ increases $r_H$, the lower bound pf \ref{sup35} increases with increasing value of the cosmological constant, while $E_+$ and $a$ are held fixed.

 Finally, taking once again the customary value $E_+=1$, one obtains for the extremal Kerr black hole : ${\bf |v|}\geq 0.414$, this yields from \ref{sup31}, ${\bf| v_{\rm rel}|}\geq 0.707$.  On the other hand, taking the triply degenerate extremal limit of KdS once again, \ref{a7}, we find ${\bf |v|}\geq 0.529$.  This yields   ${\bf| v_{\rm rel}|}\geq 0.826 $. 
%%%%%%%%%%%%%%%%%%%%%%%%%%%%%%%%%%%%%%%%%%%%%%%%%%%%%%%%
\section{Generalized area theorem for Kerr-de Sitter}\label{sec.5}
%%%%%%%%%%%%%%%%%%%%%%%
So far we have established two things : a) a positive $\Lambda$ weakens energy extraction via the Penrose process and b) energy extraction in the cosmological ergoregion is never possible. We shall now prove below that  both the horizon areas increase under this process,  thereby providing a physical evidence in favour of the generalized area theorem for rotating de Sitter black holes. This is not at all obvious {\it a priori}, owing to the results summarized at the end of \ref{sec.2}. Precisely, we have seen that the increase in the parameter $M$ increases $r_H$ but decreases $r_C$ whereas increase in the parameter $a$ does the opposite. Now, when an ejecta (carrying positive energy and angular momentum) comes out of the black hole ergoregion and escapes through the cosmological event horizon, it is clear that the parameters $M$ and $a$ describing 
the spacetime metric in the region bounded by the two horizon get decreased, yielding separately, opposite effects on $r_H$ and $r_C$.  Also, the Smarr formula for de Sitter black holes reads~\cite{Gibbons:1977mu},
$$\int_{\Sigma} \delta T_{ab} (\partial_t)^a d\Sigma^b =- \frac{\kappa_C \delta A_C +\kappa_H \delta A_H}{8\pi}-\Omega_H^{\rm Rel} \delta J_H  $$
where $\Omega_H^{\rm Rel}$ denotes the relative angular speed of the black hole horizon with respect to the cosmological horizon, $J_H$ is the angular momentum of the black hole and $\Sigma$ denotes a spacelike hypersurface located in between the two horizons; the surface gravities $\kappa_H$, $\kappa_C$
are given by \ref{sup10}. For the Penrose process in particular, one may expect the left hand side to be negative due to the energy extraction process. Likewise we should have $\delta J_H <0$. Thus it is not obvious {\it a priori} whether under this process both $A_H$ and $A_C$ or at least the sum of them would indeed increase.

Hence it needs to be proven that under this process, nevertheless, the area theorem is preserved. There is another non-triviality pertaining the definition of the mass in this case, as we shall see below.  But before we go into that, let us first briefly recall how the area theorem in this process is established in the Kerr spacetime~\cite{ Wald:1984rg, Chandrasekhar:1985kt, Christodoulou:1970wf}. 

Since the horizon Killing vector field $\chi_H^a$, \ref{sup10k}, is future directed null on BEH, we must have $-u_a\chi^a_H\geq 0$ there, giving~\cite{Wald:1984rg},
\begin{eqnarray}
E-L\Omega_H\geq 0
\label{sup36'}
\end{eqnarray}
The energy $E$ and the angular momentum $L$ of the particle changes respectively, the mass $M$ and the total angular momentum $J=Ma$ of the black hole. We have from the above equation 
\begin{eqnarray}
\left(r_H^2+a^2\right)\delta M \geq a\delta J
\label{sup36}
\end{eqnarray}
where $\delta J=a\delta M+M\delta a $. The above inequality then can be rewritten as 
\begin{eqnarray}
\left(r_H^2\delta M-Ma\delta a\right) \geq 0
\label{sup37}
\end{eqnarray}
For the Kerr black hole,  $r_H=M+\sqrt{M^2-a^2}$ and the horizon area is given by $A_H=4\pi (r_H^2+a^2)$, which is rewritten as, 
\begin{eqnarray}
A_H= 8\pi \left(M^2+M\sqrt{M^2-a^2}\right)
\label{sup38}
\end{eqnarray}
Taking the first order variation of the above equation, we find after a little algebra
\begin{eqnarray}
\delta A_H= \frac{16\pi}{\sqrt{M^2-a^2}} \left(r_H^2\delta M-Ma\delta a\right)\geq 0
\label{sup39}
\end{eqnarray}
where the inequality follows from \ref{sup37}. This proves that in the Penrose process the black hole horizon area always increases. On the other hand since $\partial A_H/\partial a \leq 0$ and $\partial A_H/\partial M \geq 0$, \ref{sup38}, and also 
$A_H$ has its greatest value for $a=0$ and the smallest value with $M=a$, it is clear that under the Penrose process, the reduction of the black hole mass is slower in rate than the reduction in the angular momentum.  A Kerr black hole thus would eventually evolve towards less spinning states and starting from the extremal limit $a=M$ we cannot reach a naked curvature singularity for which $a>M$. From \ref{sup38} and the fact that $\delta A_H\geq 0$, we can define a irreducible mass $M_{\rm irr}$~\cite{Christodoulou:1970wf},     
\begin{eqnarray}
M_{\rm irr}^2=\frac{1}{2}M\left(M+\sqrt{M^2-a^2} \right)
\label{sup40}
\end{eqnarray}
which effectively increases in the Penrose  process.\\

\noindent
Let us come back to our focus -- the KdS spacetime. We first note in this case that we cannot simply interpret  the mass parameter $M$  to be the mass or energy, as follows. Within  BEH, apart from the mass $M$ itself, there should be local positive energy due to the cosmological constant as well. Second, as we have also mentioned earlier, the rotation of the black hole  induces rotational kinetic energy onto the cosmological constant due to the frame dragging, as is evident from the $H_0^2a^2$ term appearing in various metric functions. Thus since any change in $M$  and  $a$   due to the infall of a particle changes the horizon size too, we must take into account the changes in the aforementioned local energies associated with the cosmological constant as well. 

Indeed, for the Schwarzschild-de Sitter black hole, one can have a local, Tolman-like mass function~\cite{Bhattacharya:2013tq} (see also~\cite{Dolan:2013ft}),
$$M_{\rm loc}(r)=M+\frac{H_0^2 r^3}{2}$$
in terms of which the metric function reads
$$ds^2=-(1-2M_{\rm loc}(r)/r)dt^2+(1-2M_{\rm loc}(r)/r)^{-1}dr^2+r^2d\Omega^2$$

For KdS, we {\it define}  the following local, continuous,   positive definite effective mass function 
\begin{eqnarray}
M_{\rm loc}(r):= M+\frac{H_0^2 r}{2}\left(r^2+a^2\right)
\label{sup41}
\end{eqnarray}
For $a=0$, the mass function reduces to that of the static case  whereas putting $\Lambda=0$ recovers the standard mass parameter of the Kerr spacetime. Clearly, the above mass function takes care of the spacetime rotation.  In terms of $M_{\rm loc}(r)$, the function  $\Delta_r$ in \ref{sup2} can be  written as 
$$\Delta_r=r^2-2M_{\rm loc}(r) r+a^2$$
In order to further justify this choice, let us note that intuitively, the size of a black hole should always be expected to increase with the increase of the total mass or energy contained within it.   Indeed, on the horizon $\Delta_r=0$, we can write using $M_{\rm loc} (r)$ a transcendental equation,
\begin{eqnarray}
r_H=M_{\rm loc, H} +\sqrt{M_{\rm loc,H}^2-a^2}
\label{sup42}
\end{eqnarray}
where $M_{\rm loc, H}$ is the value of $M_{\rm loc}(r)$ on $r=r_H$. It is very easy to see from the above equation that  that $\partial _{M_{\rm loc, H} }r_H \geq 0$.

Having successfully identified the mass function, the rest becomes straightforward. \ref{sup36} now takes the form,
\begin{eqnarray}
\left(r_H^2+a^2\right)\delta M_{\rm loc,H} \geq a\delta J_{\rm loc, H}
\label{sup43}
\end{eqnarray}
with $J_{\rm loc, H}:=aM_{\rm loc, H}$ and  $\delta J=a\delta M_{\rm loc,H}+M_{\rm loc,H}\delta a $. Thus we have
\begin{eqnarray}
\left(r_H^2\delta M_{\rm loc,H}-M_{\rm loc,H}a\delta a\right) \geq 0
\label{sup44}
\end{eqnarray}
The variation of the black hole horizon area is found from \ref{sup10'},
\begin{eqnarray}
\delta A_H = 4\pi\frac{\delta (r_H^2+a^2)}{1+H_0^2a^2} -8\pi \frac{(r_H^2+a^2)H_0^2 a\delta a}{(1+H_0^2a^2)^2}
\label{sup45}
\end{eqnarray}
Using the transcendental equation \ref{sup42} we can write
$$(r_H^2+a^2)=2\left(M_{\rm loc,H}^2+M_{\rm loc,H}\sqrt{M_{\rm loc,H}^2-a^2}\right)$$
Linear variation of the above equation and use of \ref{sup44} gives that the first term on the right hand side of \ref{sup45} is always greater than or equal to zero. On the other hand, since we have seen in \ref{sec.3} that the negative energy particle entering the black hole must have negative angular momentum, we must have $\delta a <0$ above, showing $\delta A_H\geq 0$.

Let us now imagine that the particle moving outward carrying positive energy and angular momentum  reaches the CEH and escapes through it. How does the area of the CEH change? We must use on the CEH,
$$E-\lambda \Omega_C \geq 0$$
We also need to work with \ref{sup41} evaluated at $r=r_C$,
\begin{eqnarray}
M_{\rm loc, C}:= M+\frac{H_0^2 r_C}{2}\left(r_C^2+a^2\right) \qquad J_{\rm loc, C}:=a M_{\rm loc, C}
\label{sup41'}
\end{eqnarray}
We follow procedure exactly similar to that of the BEH. Recalling that $\delta a <0$ in this case as well (i.e., some positive angular momentum is going out off the region bounded by the CEH),
we can show that $\delta A_C \geq 0$. This proves the generalized area theorem or the second law of thermodynamics under the Penrose process in the Kerr-de Sitter spacetime. To the best of our knowledge, this is the first explicit demonstration of that  theorem for rotating black holes in de Sitter spacetime. 

Finally, the irreducible mass function in \ref{sup40} now takes two local values
\begin{eqnarray}
M_{\rm irr;H,C}^2=\frac{1}{2}M_{\rm loc, H,C}\left(M_{\rm loc, H,C}+\sqrt{M_{\rm loc, H,C}^2-a^2} \right)
\label{sup40'}
\end{eqnarray}
 Let us now  look at \ref{a17} with $\delta H_0=0$,
\begin{eqnarray}
\Delta_r(a+\delta a,\, M+\delta M)=\Delta_r( a, M) +2a (1-H_0^2r^2)\,\delta a-2r\delta M
\label{sup46-}
\end{eqnarray}
 In the earlier location of the horizons, $r_H,\,r_C$  (corresponding to $\Delta_r(a,M)=0$), we have 
\begin{eqnarray}
\Delta_r(a+\delta a,\, M+\delta M)\vert_{r=r_H,\,r_C}=2\left[a (1-H_0^2r_{H,C}^2)\,\delta a-r_{H,C}\delta M\right]
\label{sup46}
\end{eqnarray}
 For the Penrose process  both $\delta a$ and $\delta M$ are negative. Now let us consider the increase in the black hole event horizon in this process, such that the earlier horizon radius $r_H$ is now located within the new horizon radius. This means the left hand side of the above equation is negative. Then since $a< r_H$ and $(1-H_0^2r^2)<1$ for KdS (cf. \ref{sec.2}),  it is clear that the change in the rotation parameter $a$ should be faster than the change in the mass parameter $M$. On the other hand since the cosmological horizon also increases when the outgoing ejecta crosses it, we must have the left hand side of \ref{sup46} positive on the earlier location of the CEH. Putting these all in together, we have
\begin{eqnarray}
\frac{a (1-H_0^2r_{C}^2)}{r_C}\leq \left(\frac{\delta M}{\delta a}\right)_{\rm KdS}\leq \frac{a (1-H_0^2r_{H}^2)}{r_H}
\label{sup47}
\end{eqnarray}
 Since the rotation parameter decreases faster than the mass parameter, under the Penrose process KdS spacetime evolves to the Schwarzschild-de Sitter. Furthermore, from the area theorem we proved above along with the fact that the empty de Sitter horizon
 area ($4\pi/H_0^2 $) is greater than  the sum of the two horizon areas of the Schwarzschild-de Sitter (see e.g.~\cite{Maeda:1997fh}),    leads to the same upper bound on the  total horizon area of KdS,
\begin{eqnarray}
(A_H+A_C)_{\rm KdS}  < \frac{4\pi}{H_0^2}
\label{sup48}
\end{eqnarray}
Finally, it is also clear that if we consider the Penrose process in the triply degenerate limit discussed in \ref{B}, it would spin down making all the ratios $a/M$, $aH_0$ (also $MH_0$)  (\ref{a5'}, \ref{a6}) smaller and would move away from the extremal point instead of creating a naked curvature singularity.

%%%%%%%%%%%%%%%%%%%%%%%%%%%%%%%%%%%%%%%
\section{Summary and outlook}\label{sec.6}
%%%%%%%%%%%%%%%%%%%%%%%%%%
In this work we have established three chief results :  a) a positive $\Lambda$ reduces the efficiency of the Penrose process b) the Penrose process is never possible in the cosmological ergoregion and c)  the generalized area theorem, or the second law of the de Sitter black hole mechanics  is satisfied in this process.  Note that when one deals with the classical energy-momentum tensor $T_{ab}$ of a matter field, the area theorem for a Killing horizon is satisfied if $T_{ab}$ obeys the null  or the strong energy condition, which essentially guarantees that the energy current corresponding to $T_{ab}$ along any future directed null vector is non-spacelike and future directed  on the horizon~\cite{Wald:1984rg}.  Hence the causality conditions, $-u_a\chi^a_{H,C}\geq 0$, we used on the horizons may be interpreted as the analogues of such energy conditions, for the case of a single particle picture we were concerned with.   We have pointed out in relevant places, the  qualitatively new local effects due to $\Lambda$, induced from the frame dragging. We also have derived various conclusions on the relative horizon sizes of rotating and non-rotating de Sitter black holes (\ref{sec.2}) and an upper bound the total horizon area of the KdS, \ref{sup48}. The inclusion of electric charge in our analysis should be absolutely straightforward.

Several new things could be investigated as follow up works. For example, a systematic analysis of the superradiance for various spin fields (see~\cite{Tachizawa:1992ue, Anninos:2010gh, Zhang:2014kna,   Georgescu:2014yea, Zhu:2014sya} for scalar) in KdS and evaluating their profile in various extremal limits would be highly interesting. Also, it seems very important to establish a topological version of the generalized area theorem for de Sitter black holes, analogous to that of the asymptotic flat spacetimes e.g.~\cite{Wald:1984rg}, and references therein.  Apart from this, as we have discussed in \ref{B}, the triply degenerate extremal KdS solution, just like the doubly degenerate  Nariai limit~\cite{ Anninos:2010gh}, might offer interesting quantum field theories. We shall come back to these issues in our future publications.

%%%%%%%%%%%%%%%%%%%%%%%%%%%%%%%%%%%%%%%%%%%%%%%%%%%%%%%%%%%%%%%%%%%%%%%%%%%%%%%%%%%%%%
\section*{Acknowledgement} I sincerely acknowledge  T.~Padmanabhan for various suggestions and valuable discussions, in particular, regarding the cosmological ergosphere. I would also like to thank N.~K.~Dadhich, K.~Lochan and S.~Chakraborty for useful discussions. I acknowledge anonymous referees for useful comments.
% and  Md Sabir Ali for his help regarding the diagram. 
%%%%%%%%%%%%%%%%%%%%%%%%%%%%%%%%%%%%%%%%%%%%%%%%%%%%%%%%%%%%%%%%%%%%%%%%%%%%%%%%%%%%%
\bigskip
\appendix
\labelformat{section}{Appendix #1} 
%%%%%%%%%%%%%%%%%%%%%%%%%%%%%%%%%%%%%%%%%%%%%%%%%%%%%%%%%%%%%%%%%%%%%%%%%%%%%%%%%%%%%%%%%%%%%%%%%%%
%%%%%%%%%%%%%%%%%%%%%%%%%%%%%%%%%%%%%%%%%%%%%%%%%
\section{Comparative horizon sizes of empty de Sitter, the static and rotating de Sitter black holes}\label{A}
%%%%%%%%%%%%%%%%%%%%%%%%%%%%%%%
We first note the different signs  of the function  $\Delta_r$ in \ref{sup2} in different regions,
\begin{eqnarray}
\Delta_r(r_H < r < r_C)>0,\quad \Delta_r(r_H,r_C, r_-)=0\quad \Delta_r(r_-<r<r_H)<0 \quad  \Delta_r(r<r_-)>0 ~{\rm and}~ \Delta_r(r>r_C)<0 \nonumber\\
\label{a14'}
\end{eqnarray}
We write
\begin{eqnarray}
\Delta_r=-H_0^2 \left[r^2\left(r^2-\frac{1}{H_0^2}\right)+a^2 \left(r^2-\frac{1}{H_0^2}\right) +\frac{2Mr}{H_0^2}\right]
\label{a14}
\end{eqnarray}
For the empty de Sitter spacetime, the CEH is located at $r=H_{0}^{-1}$. Thus 
\begin{eqnarray}
\Delta_r(r=H_0^{-1})=-\frac{2M}{H_0}<0
\label{a15}
\end{eqnarray}
which is only possible when $r=H_0^{-1}$ is located outside the CEH of KdS, \ref{a14'}.
Thus the CEH of KdS is smaller in size than the empty de Sitter. Likewise the horizons of the Schwarzschild-de Sitter ($a=0$) spacetime is given by the zeros of the function
\begin{eqnarray}
-\left[r\left(r^2-\frac{1}{H_0^2}\right)+ \frac{2M}{H_0^2} \right]
\label{a15'}
\end{eqnarray}
On the location of the CEH of the empty de Sitter ($=H_0^{-1}$), the above function  is negative. It is negative   also at $r=2M$, the horizon radius of the Schwarzschild spacetime. This shows that a) the CEH of the Schwarzschild-de Sitter is smaller than that of the empty de Sitter whereas b) the BEH of the of the same is larger than that of a Schwarzschild black hole with the same mass parameter.

On the locations of the Schwarzschild-de Sitter horizons (\ref{a15'}$=0$), we have for the KdS, 
\begin{eqnarray}
\Delta_r=-H_0^2a^2 \left(r^2-\frac{1}{H_0^2}\right)
\label{a16}
\end{eqnarray}
which is  positive because as we have seen, for both KdS and the Schwarzschild-de Sitter, the cosmological horizon scale is smaller than that of the empty de Sitter ($H_0^{-1}$). Since $\Delta_r >0$ for KdS at the Schwarzschild-de Sitter horizon lengths, it is clear that the BEH in the KdS is smaller than that of the Schwarzschild-de Sitter, whereas the CEH is larger than the same (the parameter $M$ held fixed). 

Putting these all in together, perhaps we could generically conjecture that : a) the increase in $M$ of a black hole increases its horizon size but decreases the size of the CEH b) the increase in the cosmological constant increases the black hole size but does the opposite to the CEH c) the increase in rotation decreases the size of the black hole but increases the size of the CEH. Let us see these explicitly by making an infinitesimal change in the parameters : $a\to a+\delta a$, $M\to M+\delta M$ and $H_0 \to H_0+\delta H_0$. We have at the linear order,
\begin{eqnarray}
\Delta_r(a+\delta a,\, M+\delta M, H_0+\delta H_0)=\Delta_r( a, M, H_0)-2H_0r^2 (r^2+a^2)\delta H_0 +2a (1-H_0^2r^2)\,\delta a-2r\delta M
\label{a17}
\end{eqnarray}
and evaluate the sign of the left hand side at the old horizon points, $\Delta_r( a, M, H_0)=0$. If we take $\delta M \neq 0$ only, we have $\Delta_r(a+\delta a,\, M+\delta M, H_0+\delta H_0)<0$ on the old horizons' locations. This implies that for $\delta M>0$ size of the BEH/CEH has increased/decreased, \ref{a14'}. Likewise for $\delta a \neq 0$ only, the size of the    BEH/CEH decreases/increases for $\delta a >0$, follows from our earlier result that the CEH size in the KdS is always smaller than that of the empty de Sitter. Finally, for $\delta H_0 >0$ the size of the BEH/CEH increases/decreases. Thus, since the minimum horizon radius in the Kerr spacetime is  either $M$ or $a$, we must have for the KdS the lower bounds : $r_H>M$ or $r_H>a$.

%%%%%%%%%%%%%%%%%%%%%%%%%%%%%%%%%%%%%%%%%%%%%%%%%%%%%%%%%%%%%%%%%%%%%%%%%%%%%%%%%%%%%%%%%%%%%%%%%%%
%%%%%%%%%%%%%%%%%%%%%%%%%%%%%%%%%%%%%%%%%%%%%%%%%%%%%%%%%%%%%%%%%%%%%%%%%%%%%%%%%%%%%%%%%%%%%%%%%%%
%%%%%%%%%%%%%%%%%%%%%%%%%%%%%%%%%%%%%%%%%%%%%%%%%%%%%%%%%%%%%%%%%%%%%%%%%%%%%%
\section{The triply degenerate  limit of the Kerr-de Sitter}\label{B}
\noindent 
We wish to derive a triply degenerate extremal limit of the Kerr-de Sitter spacetime. For a doubly degenerate or the Nariai limit, we refer our reader to~\cite{Anninos:2010gh}. We have to solve $\Delta_r=0$ or equivalently,
\begin{eqnarray}
r^4-\frac{(1-H_0^2a^2)r^2}{H_0^2}+\frac{2Mr}{H_0^2}-\frac{a^2}{H_0^2}=0
\label{a1}
\end{eqnarray}
if $r_i\,(i=1,2,3,4)$ are the four roots of this equation, we have
\begin{eqnarray}
&&r_1+r_2+r_3+r_4=0 \qquad r_1r_2r_3r_4=-\frac{a^2}{H_0^2} \nonumber\\
&&r_1r_2+r_2r_3+r_3r_4+r_4r_1+r_3r_1+r_2r_4=-\frac{(1-H_0^2a^2)}{H_0^2}\nonumber\\
&&r_1r_2r_3+r_2r_3r_4+r_3r_4r_1+r_4r_1r_2=-\frac{2M}{H_0^2}
\label{a2}
\end{eqnarray}
The first of the above equations shows that not all the four roots can be simultaneously real and positive.  We wish to have three positive real roots which, in their order of increasing sizes, are respectively the inner or the Cauchy horizon, the black hole event horizon and the cosmological event horizon. Then taking $r_4<0$ we get by using the first and second of the above equations,  
\begin{eqnarray}
r_1^2+r_2^2+r_3^2+r_1r_2+r_2r_3+r_3r_1=\frac{(1-H_0^2a^2)}{H_0^2}
\label{a3}
\end{eqnarray}
Since the left hand side is positive definite, we  immediately conclude 
\begin{eqnarray}
a < \frac{1}{H_0}
\label{a4}
\end{eqnarray}
We are chiefly interested here in finding analytically the triply degenerate situation. Writing  $r_1=r_2=r_3=r_0$, it is easy to obtain from \ref{a2},
\begin{eqnarray}
r_0=\left(\frac{a^2}{3H_0^2}\right)^{1/4}=\frac{(1-H_0^2a^2)^{1/2}}{\sqrt{6}H_0}=\left(\frac{M}{4H_0^2}\right)^{1/3}
\label{a5}
\end{eqnarray}
This shows that the root is expressible via only one parameter -- either $M$ or $H_0$ or $a$. For instance, equating the first two and using \ref{a4}, we obtain
\begin{eqnarray}
aH_0=2-\sqrt{3} 
\label{a5'}
\end{eqnarray}
which is consistent with \ref{a4}. Likewise 
\begin{eqnarray}
MH_0=\frac{4}{3^{3/4}} \left(2-\sqrt{3} \right)^{3/2}~~{\rm and}~~~ \frac{a}{M}=\frac{3^{3/4}}{4(2-\sqrt{3})^{1/2}}\approx 1.01
\label{a6}
\end{eqnarray}
showing $a=M$ is not the extremal is not the extremal limit of the KdS, pointed out earlier in~\cite{Lake:2015xca} by analysis of the parameter space of he KdS spacetime. We can re-express the horizon length as
\begin{eqnarray}
r_0=\frac{\sqrt{3} M}{4(2-\sqrt{3})}=\frac{\left(2-\sqrt{3}\right)^{1/2}}{3^{1/4} H_0}=\frac{a}{3^{1/4}(2-\sqrt{3})^{1/2}}\approx 1.62 M
\label{a7}
\end{eqnarray}
If we have a charged black hole, the rotation parameter simply gets replaced with $\sqrt{a^2+Q^2}$. We can compare \ref{a6} with the known cases. For the extremal Kerr black hole ($H_0=0$), we must have $a/M=1$. Also, for the static Nariai class of black holes $(a=0)$ we have $3M\sqrt{\Lambda}=1\equiv MH_0=0.192$. 

We shall now also prove that the limits of \ref{a5'}, \ref{a6} are really extreme for if we try to increase the value of any of the parameters $M,\,a $ and $H_0$, we create a naked curvature singularity, following \ref{A}. First, recall that we have already proven increasing $H_0$ or $M$ would increase the BEH size but decrease the CEH size. Thus when they are coincident, increasing either or both these parameters would certainly destroy both of them. What happens when we increase $a$? Let us recall that increasing $a$ would decrease the BEH and increase the CEH, leading to no definite conclusion. However, note that the function $\Delta_r$ (\ref{a14}) is negative in the region between the BEH and the Cauchy horizon. Then following exactly the same procedure  as in \ref{A}, we can show that increasing $a$ actually increases the Cauchy horizon size. Since the BEH and the Cauchy horizon behaves oppositely for $\delta a>0$, certainly in the coincident limit, increasing $a$ any further would destroy the horizon structure. In other words, in  a regular KdS spacetime, we must have the limits of \ref{a5'}, \ref{a6} or \ref{a7'} satisfied. We can also compare this triply degenerate solution once again to the doubly degenerate Nariai one derived in~\cite{Anninos:2010gh}, where only the BEH and the CEH are coincident. Note that for such a case, we may definitely increase $a$, making the CEH bigger and the BEH smaller, until the BEH gets coincident with the Cauchy horizon.  \\

\noindent
Let us now also obtain a solution for the Reissner-N\"{o}rdstrom-de Sitter solution with triply degenerate Killing horizons. The metric reads
\begin{eqnarray}
ds^2=-\left(1-\frac{2M}{r}-H_0^2r^2+\frac{Q^2}{r^2}\right)dt^2+\left(1-\frac{2M}{r}-H_0^2r^2+\frac{Q^2}{r^2}\right)^{-1}dr^2+r^2d\Omega^2
\label{a8}
\end{eqnarray}
Like the KdS, here also the horizons are determined by the roots of a quartic equation,
\begin{eqnarray}
r^4-\frac{r^2}{H_0^2}+\frac{2Mr}{H_0^2}-\frac{Q^2}{H_0^2}=0
\label{a9}
\end{eqnarray}
The analogue of \ref{a2} now becomes 
\begin{eqnarray}
&&r_1+r_2+r_3+r_4=0 \qquad r_1r_2r_3r_4=-\frac{Q^2}{H_0^2} \nonumber\\
&&r_1r_2+r_2r_3+r_3r_4+r_4r_1+r_3r_1+r_2r_4=-\frac{1}{H_0^2}  \nonumber\\
&&r_1r_2r_3+r_2r_3r_4+r_3r_4r_1+r_4r_1r_2=-\frac{2M}{H_0^2}
\label{a10}
\end{eqnarray} 
We find for the triply degenerate case,
\begin{eqnarray}
r_0=\left(\frac{Q^2}{3H_0^2}\right)^{1/4}=\frac{1}{\sqrt{6}H_0}=\left(\frac{M}{4H_0^2}\right)^{1/3}
\label{a11}
\end{eqnarray}
with the conditions
\begin{eqnarray}
QH_0=\frac{1}{2\sqrt{3}}, \qquad MH_0=\frac{\sqrt{2}}{3\sqrt{3}}, \qquad \frac{Q}{M}= \frac{3}{2\sqrt{2}}
\label{a12}
\end{eqnarray}
Note  the chief difference between the known extremal black holes with the ones we have demonstrated -- {\it all} known extremal black holes have doubly degenerate Killing horizons, whereas here we have {\it triply degenerate} ones, indicating 
novel geometry and quantum field theory in the vicinity of such horizons. Hopefully such black holes can be important in the early universe or in the context of the false vacuum decay scenario.

Certainly,  such triply degenerate solutions are qualitatively  different from their doubly degenerate Nariai counterparts, where the black hole and the cosmological event horizons are nearly coincident. For example, for the Resissner-N\"{o}rdstrom-de Sitter spacetime, we can find such a Nariai solution for $Q=M$  and $4M H_0=1$, for which the black hole and cosmological and the Cauchy horizons are respectively located at
\begin{eqnarray}
r_H=\frac{1}{2H_0}=r_C \qquad r_{-}=\frac{\sqrt{2}-1}{2H_0}
\label{a13}
\end{eqnarray}
%

%%%%%%%%%%%%%%%%%%%%%%%%%%%%%%%%%%%%%%%%%%%%%%%%%%%%%%%%%%%%%%%%%%%%%%%%%%%%%
%%%%%%%%%%%%%%%%%%%%%%%%%%%%%%%%%%%%%%%%%%%%%%%%%%%%%%%%%%%%%%%%%%%%%%%%%%%%%%
\section{A proof that Carter's constant $\geq 0$ }\label{C}
%%%%%%%%%%%%%%%%%%%%%%%%%%%%
If the motion is confined to the equatorial plane $\theta =\pi/2$ only, clearly we have $\zeta=0$ then. Now on the right hand side of \ref{sup7}, all but the third term are negative. Let us then consider the function
$$\frac{\sin^2\theta}{\Delta_{\theta}}\left(aE-\frac{\Xi L}{\sin^2\theta}\right)^2-(\Xi L-aE)^2$$
Clearly the first term dominates as we move towards the axis, $\theta=0, \pi$. On the other hand on $\theta=\pi/2$ they become equal. Let us now check whether there exists any point in the intervals  $0<\theta<\pi/2$ or $\pi/2 < \theta <\pi$, where the second term dominates the first. If this is indeed the case, we shall  have at least one zero of the above function in the above intervals and hence at least one extremum. However, it is easy to check that the only extremum of the above function is located at $\theta=\pi/2$, proving that the above function remains always greater than or equal to zero in the entire domain $0\leq \theta \leq \pi$. Thus the right hand side of \ref{sup7} is negative, proving $\zeta \geq 0$, {\it always}.

\bigskip

%%%%%%%%%%%%%%%%%%%%%%%%%%%%%%%%%%%%%%%%%%%%%%%%%%%%%%%%%%%%%%%%%%%
%\section*{Acknowledgements}
%\noindent

%%%%	```````%%%%%%%%%%%%%%%%%%%%%%%%%%%%%%%%%%%%%%%%%%%%%%%%%%%%%%%%%%%%%%%%%%%%%%%


\begin{thebibliography}{99} 

%%%GEN REFs
   
\bibitem{Wald:1984rg}
R.~M.~Wald, ``{\it General Relativity}'', Chicago Univ. Press (1986).
 
 
 \bibitem{Chandrasekhar:1985kt} 
  S.~Chandrasekhar,
  ``{\it The mathematical theory of black holes}'',
  Oxford, UK: Clarendon (1985).
  

  
  \bibitem{Paddy}
   T.~Padmanabhan, ``{\it Gravitation : Foundation and Frontiers}'', Cambridge Univ. Press (2010).
 
   
   
 \bibitem{Wald}
R.~M.~Wald,  ``{\it Gedanken Experiments to Destroy a Black Hole}'',
 Ann.\ Phys. {\bf 83}, 548 (1974).
 
  
  
  
  
 
 \bibitem{Christodoulou:1970wf} 
  D.~Christodoulou,
  ``{\it Reversible and irreversible transforations in black hole physics}'',
  Phys.\ Rev.\ Lett.\  {\bf 25}, 1596 (1970).
 
 \bibitem{Wald:1974kya} 
  R.~M.~Wald,
  ``{\it Energy Limits on the Penrose Process}'',
  Astrophys.\ J.\  {\bf 191}, 231 (1974).
  
  
  \bibitem{Bardeen:1972fi} 
    J.~M.~Bardeen, W.~H.~Press and S.~A.~Teukolsky,
  ``{\it Rotating black holes: Locally nonrotating frames, energy extraction, and scalar synchrotron radiation}'',
  Astrophys.\ J.\  {\bf 178}, 347 (1972).
  
  \bibitem{Christodoulou:1972kt} 
  D.~Christodoulou and R.~Ruffini,
  ``{\it Reversible transformations of a charged black hole}'',
  Phys.\ Rev.\ D {\bf 4}, 3552 (1971).
 
  \bibitem{Wagh:1986tsa} 
  S.~M.~Wagh, S.~V.~Dhurandhar and N.~Dadhich,
  ``{\it Revival of penrose process for astrophysical applications}",
    Astrophys.\ J.\  {\bf 290}, no. 12 (1985)
  [{\it Erratum :} Astrophys.\ J.\  {\bf 301}, 1018 (1986)].
  
   \bibitem{Dadhich:2012yu} 
  N.~Dadhich,
  ``{\it Magnetic Penrose Process and Blanford-Zanejk mechanism: A clarification}'',
  arXiv:1210.1041 [astro-ph.HE].
 

 
  %%%%%%%%%%%%%% SUP in Asymp. flat space
 
 \bibitem{Leite:2017zyb}  
  L.~C.~S.~Leite, S.~R.~Dolan and L.~C.~B.~Crispino,
  ``{\it Absorption of electromagnetic and gravitational waves by Kerr black holes: Shadows, superradiance and the spin-helicity effect}'',
  arXiv:1707.01144 [gr-qc].
 
 \bibitem{East:2017mrj} 
  W.~E.~East,
  ``{\it Superradiant instability of massive vector fields around spinning black holes in the relativistic regime}'',
  Phys.\ Rev.\ D {\bf 96}, no. 2, 024004 (2017)
  [arXiv:1705.01544 [gr-qc]].
  
  
  \bibitem{Cardoso:2017kgn} 
  V.~Cardoso, P.~Pani and T.~T.~Yu,
  ``{\it Superradiance in rotating stars and pulsar-timing constraints on dark photons}'',
  Phys.\ Rev.\ D {\bf 95}, no. 12, 124056 (2017)
  [arXiv:1704.06151 [gr-qc]].
  
  
  \bibitem{Hod:2017cga} 
  S.~Hod,
  ``{\it Onset of superradiant instabilities in rotating spacetimes of exotic compact objects}'',
  JHEP {\bf 1706}, 132 (2017)
  [arXiv:1704.05856 [hep-th]].
  
  
 \bibitem{East:2017ovw} 
  W.~E.~East and F.~Pretorius,
  ``{\it Superradiant Instability and Backreaction of Massive Vector Fields around Kerr Black Holes}'',
  Phys.\ Rev.\ Lett.\  {\bf 119}, no. 4, 041101 (2017)
  [arXiv:1704.04791 [gr-qc]].
  
  
  \bibitem{Huang:2016qnk} 
  Y.~Huang and D.~J.~Liu,
  ``{\it Scalar clouds and the superradiant instability regime of Kerr-Newman black hole}'',
  Phys.\ Rev.\ D {\bf 94}, no. 6, 064030 (2016)
  [arXiv:1606.08913 [gr-qc]].
 
 
 \bibitem{Hod:2016iri} 
  S.~Hod,
  ``{\it The superradiant instability regime of the spinning Kerr black hole}'',
  Phys.\ Lett.\ B {\bf 758}, 181 (2016)
  [arXiv:1606.02306 [gr-qc]].
  
  
   
 \bibitem{Konoplya}
  R.~A.~Konoplya,
  ``{\it Magnetic field creates strong superradiant instability}''
  Phys.\ Lett.\ B {\bf 666}, 283 (2008)
  [Phys.\ Lett.\ B {\bf 670}, 459 (2009)]
  [arXiv:0801.0846 [hep-th]].
 
 
 \bibitem{Brito:2014nja} 
  R.~Brito, V.~Cardoso and P.~Pani,
  ``{\it Superradiant instability of black holes immersed in a magnetic field}'',
  Phys.\ Rev.\ D {\bf 89}, no. 10, 104045 (2014)
  [arXiv:1405.2098 [gr-qc]].
 
  
 
  
  \bibitem{Winstanley:2001nx} 
  E.~Winstanley,
  ``{\it On classical superradiance in Kerr-Newman - anti-de Sitter black holes}'',
  Phys.\ Rev.\ D {\bf 64}, 104010 (2001)
  [gr-qc/0106032].
  
   
 \bibitem{Aliev:2015wla} 
  A.~N.~Aliev,
  ``{\it Superradiance and instability of small rotating charged AdS black holes in all dimensions}'',
  Eur.\ Phys.\ J.\ C {\bf 76}, no. 2, 58 (2016)
  [arXiv:1503.08607 [hep-th]].
 
 
 
 \bibitem{Bosch:2016vcp} 
  P.~Bosch, S.~R.~Green and L.~Lehner,
  ``{\it Nonlinear Evolution and Final Fate of Charged Anti–de Sitter Black Hole Superradiant Instability}'',
  Phys.\ Rev.\ Lett.\  {\bf 116}, no. 14, 141102 (2016)
  [arXiv:1601.01384 [gr-qc]].

 \bibitem{Gonzalez:2017shu} 
  P.~A.~González, E.~Papantonopoulos, J.~Saavedra and Y.~Vásquez,
  ``{\it Superradiant Instability of Near Extremal and Extremal Four-Dimensional Charged Hairy Black Hole in anti-de Sitter Spacetime}'',
  Phys.\ Rev.\ D {\bf 95}, no. 6, 064046 (2017)
  [arXiv:1702.00439 [gr-qc]].
  
 
  
  \bibitem{Tachizawa:1992ue} 
       T.~Tachizawa and K.~i.~Maeda,
  ``{\it Superradiance in the Kerr-de Sitter space-time}'',
Phys.\ Lett.\ {\bf A172}, 325 (1993). 
 
  
 \bibitem{Anninos:2010gh} 
  D.~Anninos and T.~Anous,
  ``{\it A de Sitter Hoedown}'',
  JHEP {\bf 1008}, 131 (2010)
  [arXiv:1002.1717 [hep-th]].
  
   
 \bibitem{Zhang:2014kna} 
  C.~Y.~Zhang, S.~J.~Zhang and B.~Wang,
  ``{\it Superradiant instability of Kerr-de Sitter black holes in scalar-tensor theory}'',
  JHEP {\bf 1408}, 011 (2014)
  [arXiv:1405.3811 [hep-th]].
  
  
  
  \bibitem{Georgescu:2014yea} 
  V.~Georgescu, C.~Gérard and D.~Häfner,
  ``{\it Asymptotic completeness for superradiant Klein-Gordon equations and applications to the De Sitter Kerr metric}'',
  arXiv:1405.5304 [math.AP].
  
  
  \bibitem{Zhu:2014sya} 
  Z.~Zhu, S.~J.~Zhang, C.~E.~Pellicer, B.~Wang and E.~Abdalla,
  ``{\it Stability of Reissner-Nordström black hole in de Sitter background under charged scalar perturbation}'',
  Phys.\ Rev.\ D {\bf 90}, no. 4, 044042 (2014)
  Addendum: [Phys.\ Rev.\ D {\bf 90}, no. 4, 049904 (2014)]
  [arXiv:1405.4931 [hep-th]].
  
    
 \bibitem{Ganguly:2017zjf} 
  O.~Ganguly,
  ``{\it Acoustic superradiance in a slightly viscous fluid}'',
  arXiv:1705.04935 [gr-qc].


  
   
 \bibitem{Brito:2015oca} 
  R.~Brito, V.~Cardoso and P.~Pani,
  ``{\it Superradiance : Energy Extraction, Black-Hole Bombs and Implications for Astrophysics and Particle Physics}'',
  Lect.\ Notes Phys.\  {\bf 906}, pp.1 (2015)
  [arXiv:1501.06570 [gr-qc]].
  
  
 
  
 %%%%%%%%%%%%%%%%%%%%%%%%%%%%%% dS


  
  %% dS BH Refs
  
  \bibitem{Carter:1968ks} 
  B.~Carter,
  ``{\it Hamilton-Jacobi and Schrodinger separable solutions of Einstein's equations}'',
  Commun.\ Math.\ Phys.\  {\bf 10}, 280 (1968).
  
  
  
  \bibitem{Gibbons:2004uw} 
  G.~W.~Gibbons, H.~Lu, D.~N.~Page and C.~N.~Pope,
  ``{\it The General Kerr-de Sitter metrics in all dimensions}'',
  J.\ Geom.\ Phys.\  {\bf 53}, 49 (2005)
  [hep-th/0404008].
  
  
  \bibitem{Akcay:2010vt} 
  S.~Akcay and R.~A.~Matzner,
 ``{\it Kerr-de Sitter Universe}'',
  Class.\ Quant.\ Grav.\  {\bf 28}, 085012 (2011)
  [arXiv:1011.0479 [gr-qc]].
  
  
  \bibitem{Lake:2015xca} 
  K.~Lake and T.~Zannias,
  ``{\it Global structure of Kerr–de Sitter spacetimes}'',
  Phys.\ Rev.\ D {\bf 92}, no. 8, 084003 (2015)
  [arXiv:1507.08984 [gr-qc]].
  
  
  
 
  
  %%%% dS BH thermo 
  
  \bibitem{Gibbons:1977mu} 
  G.~W.~Gibbons and S.~W.~Hawking,
  ``{\it Cosmological Event Horizons, Thermodynamics, and Particle Creation}'',
  Phys.\ Rev.\ D {\bf 15}, 2738 (1977).
 
  
  \bibitem{Kastor:1993mj} 
  D.~Kastor and J.~H.~Traschen,
  ``{\it Particle production and positive energy theorems for charged black holes in De Sitter}'',
  Class.\ Quant.\ Grav.\  {\bf 13}, 2753 (1996)
  [gr-qc/9311025].

  
  \bibitem{Maeda:1997fh} 
  K.~Maeda, T.~Koike, M.~Narita and A.~Ishibashi,
  ``{\it Upper bound for entropy in asymptotically de Sitter space-time}'',
  Phys.\ Rev.\ D {\bf 57}, 3503 (1998)
  [gr-qc/9712029].
 
  
  
  \bibitem{Traschen:1999zr} 
  J.~H.~Traschen,
 ``{\it An Introduction to black hole evaporation}''
  gr-qc/0010055.
  
  
  \bibitem{Davies:2003me} 
  P.~C.~W.~Davies and T.~M.~Davis,
  ``{\it How far can the generalized second law be generalized?}'',
  Found.\ Phys.\  {\bf 32}, 1877 (2002)
  [astro-ph/0310522].
  
  
  \bibitem{Urano:2009xn} 
  M.~Urano, A.~Tomimatsu and H.~Saida,
  ``{\it Mechanical First Law of Black Hole Spacetimes with Cosmological Constant and Its Application to Schwarzschild-de Sitter Spacetime}'',
  Class.\ Quant.\ Grav.\  {\bf 26}, 105010 (2009)
  [arXiv:0903.4230 [gr-qc]].
  
  
  \bibitem{Saida:2009up} 
    H.~Saida,
  ``{\it de Sitter thermodynamics in the canonical ensemble}'',
  Prog.\ Theor.\ Phys.\  {\bf 122}, 1239 (2010)
  [arXiv:0908.3041 [gr-qc]].
  
  \bibitem{Saida:2009ss} 
  H.~Saida,
  ``{\it To what extent is the entropy-area law universal?: Multi-horizon and multi-temperature spacetime may break the entropy-area law}'',
  Prog.\ Theor.\ Phys.\  {\bf 122}, 1515 (2010)
  [arXiv:0910.2510 [gr-qc]].
  
 
  
  \bibitem{Bhattacharya:2013tq} 
    S.~Bhattacharya and A.~Lahiri,
  ``{\it Mass function and particle creation in Schwarzschild-de Sitter spacetime}'',
  Eur.\ Phys.\ J.\ C {\bf 73}, 2673 (2013)
  [arXiv:1301.4532 [gr-qc]].
  
 
 \bibitem{Dolan:2013ft} 
  B.~P.~Dolan, D.~Kastor, D.~Kubiznak, R.~B.~Mann and J.~Traschen,
  ``{\it Thermodynamic Volumes and Isoperimetric Inequalities for de Sitter Black Holes}'',
  Phys.\ Rev.\ D {\bf 87}, no. 10, 104017 (2013)
  [arXiv:1301.5926 [hep-th]].
  
  
  

  
  
 \bibitem{Dolan:2014jva} 
  B.~P.~Dolan,
  ``{\it Black holes and Boyle's law — The thermodynamics of the cosmological constant}'',
  Mod.\ Phys.\ Lett.\ A {\bf 30}, no. 03n04, 1540002 (2015)
  [arXiv:1408.4023 [gr-qc]].
 
  
  
  \bibitem{Li:2016zca} 
  H.~F.~Li, M.~S.~Ma and Y.~Q.~Ma,
  ``{\it Thermodynamic properties of black holes in de Sitter space}'',
  Mod.\ Phys.\ Lett.\ A {\bf 32}, no. 02, 1750017 (2016)
  [arXiv:1605.08225 [hep-th]].
  
 
  
 
   \bibitem{Altamirano:2014tva} 
  N.~Altamirano, D.~Kubiznak, R.~B.~Mann and Z.~Sherkatghanad,
  ``{\it Thermodynamics of rotating black holes and black rings: phase transitions and thermodynamic volume}'',
  Galaxies {\bf 2}, 89 (2014)
  [arXiv:1401.2586 [hep-th]].
  
   \bibitem{Zhang:2014jfa} 
    L.~C.~Zhang, M.~S.~Ma, H.~H.~Zhao and R.~Zhao,
  ``{\it Thermodynamics of phase transition in higher-dimensional Reissner-Nordström-de Sitter black hole}'',
  Eur.\ Phys.\ J.\ C {\bf 74}, no. 9, 3052 (2014)
  [arXiv:1403.2151 [gr-qc]].
  
  
  
 \bibitem{Kubiznak:2015bya} 
  D.~Kubiznak and F.~Simovic,
  ``{\it Thermodynamics of horizons: de Sitter black holes and reentrant phase transitions}'',
  Class.\ Quant.\ Grav.\  {\bf 33}, no. 24, 245001 (2016)
  [arXiv:1507.08630 [hep-th]].
  
 \bibitem{Bhattacharya:2015mja} 
  S.~Bhattacharya,
  ``{\it A note on entropy of de Sitter black holes}'',
  Eur.\ Phys.\ J.\ C {\bf 76}, no. 3, 112 (2016)
  [arXiv:1506.07809 [gr-qc]].
 
 \bibitem{Kanti:2017ubd} 
  P.~Kanti and T.~Pappas,
  ``{\it Effective temperatures and radiation spectra for a higher-dimensional Schwarzschild–de Sitter black hole}'',
  Phys.\ Rev.\ D {\bf 96}, no. 2, 024038 (2017)
  [arXiv:1705.09108 [hep-th]].
  

  
  \bibitem{Pappas:2017kam} 
    T.~Pappas and P.~Kanti,
  ``{\it Schwarzschild-de Sitter spacetime: the role of Temperature in the emission of Hawking radiation}'',
  arXiv:1707.04900 [hep-th].
 
 
   \bibitem{Bhattacharya:2017bpl} 
  S.~Bhattacharya, S.~Chakraborty and T.~Padmanabhan,
  ``{\it Entropy of a box of gas in an external gravitational field $-$ revisited}'',
  Phys.\ Rev.\ D {\bf 96}, no. 8, 084030 (2017)
  [arXiv:1702.08723 [gr-qc]].
  

 
  
   
 \bibitem{Li:2016zdi} 
  H.~F.~Li, M.~S.~Ma, L.~C.~Zhang and R.~Zhao,
  ``{\it Entropy of Kerr–de Sitter black hole}'',
  Nucl.\ Phys.\ B {\bf 920}, 211 (2017)
  [arXiv:1612.03248 [gr-qc]].
 
 
  %%%%%%%%%
 
 
%%%%%%%%%%%%%%%%%%%%%%%%%%%%%%% NEW   %%%%%%%%%%%%%%%%%%%%%%%


\end{thebibliography}
\end{document}